\documentclass[twocolumn,showpacs,preprintnumbers,amsmath,amssymb]{revtex4}
\usepackage{graphics}
\usepackage{epsfig}
\usepackage{color}
\usepackage{subfigure}  

\begin{document}

\title{CITIUS: 
\\
an IR-XUV light source for fundamental and applied ultrafast science}

\author{C. Grazioli$^{1,2}$}
\author{C. Callegari$^{2}$} 
\author{A. Ciavardini$^{3}$} 
\author{M. Coreno$^{4,2}$} 
\author{F. Frassetto$^{5}$} 
\author{D. Gauthier$^{1,2}$} 
\author{D. Golob$^{6}$} 
\author{R. Ivanov$^{1,2}$} 
\author{A. Kivim\"aki$^{7}$}
\author{B. Mahieu$^{2,8}$} 
\author{Bojan Bu\v car$^{9}$} 
\author{M. Merhar$^{9}$} 
\author{P. Miotti$^{5}$} 
\author{L. Poletto$^{5}$} 
\author{E. Polo$^{10}$} 
\author{B. Ressel$^{1}$} 
\author{C. Spezzani$^{2}$} 
\author{G. De Ninno$^{1,2}$}
\affiliation{
1. Laboratory of Quantum Optics, University of Nova Gorica, Nova Gorica, Slovenia\\
2. Elettra Sincrotrone Trieste, Trieste, Italy \\
3.	Sapienza University, Rome, Italy\\
4.	Institute of Inorganic Methodologies and Plasmas (CNR-IMIP), Montelibretti, Roma, Italy\\
5.	Institute of Photonics and Nanotechnologies (CNR-IFN), Padova, Italy\\
6.	Kontrolni Sistemi d.o.o., Se\v{z}ana, Slovenia\\
7.	Institute of Materials Manufacturing (CNR-IOM),  TASC Laboratory, Trieste, Italy\\
8.  Service des Photons Atomes et Mol\'{e}cules, Commissariat \`{a}  l'Energie Atomique, Centre d'Etudes de Saclay, B\^{a}timent 522, 91191 Gif-sur-Yvette, France\\
9.	Laboratory of Mechanical Processing Technologies, University of Ljubljana, Slovenia\\
10.	Institute of Organic Synthesis and Photoreactivity (CNR-ISOF), Ferrara, Italy\\
}
\date{\today}

\begin{abstract}

We present the main features of CITIUS, a new light source for ultrafast science, generating tunable, intense, femtosecond pulses in the spectral range from IR to XUV. The XUV pulses (about $10^5$-$10^8$ photons/pulse in the range 14-80 eV) are produced by laser-induced high-order harmonic generation in gas. This radiation is monochromatized by a time-preserving monochromator, allowing also to work with high-resolution bandwidth selection. The tunable IR-UV pulses ($10^{12}$-$10^{15}$ photons/pulse in the range 0.4-5.6 eV) are generated by an optical parametric amplifier, which is driven by a fraction of the same laser pulse that generates high order harmonics.  The IR-UV and XUV pulses follow different optical paths and are eventually recombined on the sample for pump-probe experiments. The new light source will become the fulcrum of a new center located at the University of Nova Gorica, active in a wide range of scientific fields, including materials science, catalysis, biochemistry and magnetism. We also present the results of two pump-probe experiments: with the first one, we fully characterized the temporal duration of harmonic pulses in the time-preserving configuration; with the second one, we demonstrated the possibility of using CITIUS for studying of ultra-fast dynamics.  
\end{abstract}

\maketitle

\section{Introduction}

Thanks to advances in laser science occurred over the last two decades, it is now possible to produce soft X-ray pulses in the femtosecond range, using different approaches: laser-induced high-order harmonic generation in gas (HHG), free-electron lasers, and laser slicing at synchrotron beamlines \cite{pfei}. In HHG, noble-gas atoms interact with an intense laser field and emit light at the harmonics of the latter. The process can be described by means of a semiclassical ``three step model'' 
\cite{kork}. In the first step, a bound electron is extracted from a gas atom under the action of the laser field; in the second step, the electron undergoes an oscillating motion imposed by the laser field; in the third step, the electron collides back with the parent ion. As a result, laser harmonics are emitted up to wavelengths extending into the soft X-ray regime, and droping quickly at some cutoff energy \cite{cha,spie,sere}.
\\
Since the electron dynamics in atomic, molecular or more complex systems occurs at timescales ranging from hundreds of femtoseconds down to tens of attoseconds, HHG sources provide an ideal tool to study, and possibly control, the interactions between charge, lattice, orbital, and spin dynamics in a chemical reaction.
\\
In 2008, a new project has been funded in the framework of the “Cross-border cooperation program between Italy and Slovenia 2007-2013” \cite{itaslo}. Project's main objective is to set up a state-of-the-art light source at University of Nova Gorica (Slovenia). The source, named CITIUS, has been  assembled and commissioned at Elettra-Sincrotrone Trieste (Italy) and is going to be moved to Slovenia. The mission of the new Slovenian laboratory is to carry out cutting-edge experiments in a wide range of both applied and fundamental research fields, including materials science, catalysis, biochemistry and magnetism. Part of the envisaged scientific program will be connected to that developed at the Low-Density Matter beamline at the FERMI@Elettra free-electron laser \cite{fermi}. 
\\
In this paper, we first provide a technical description of CITIUS, and then report the results of two proof-of-principle pump-probe experiments, namely the characterization of the temporal structure of HHG pulses, and the selective investigation of the ultra-fast dynamics of different elements in a magnetic compound.

\section{The laser system and the beamline}

\begin{figure}
\centering
{\resizebox{0.48\textwidth}{!}{ \includegraphics{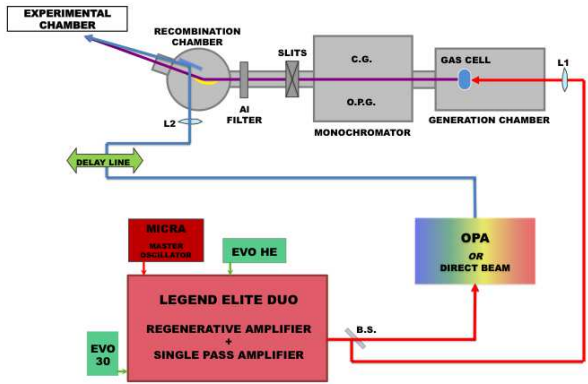}}}
\caption{Layout of the CITIUS source including the laser system, the harmonic generation chamber, the monochromator, the recombination chamber, the delay line and the experimental chamber. The light generated by the Legend amplifier is separated in two parts by the beam splitter BS. The reflected part (about two thirds of the incoming beam) is focused (with the lens L1) into the generation chamber, where it is used for HHG. The HHG beam is monochromatized, filtered (in order to eliminate the residual IR beam) and sent into the refocusing chamber. From there, it is finally focalized into the experimental chamber, using a toroidal mirror. The part of laser beam transmitted by BS is either sent to the OPA, or transmitted directly. The OPA/direct beam is then sent through a delay line, focused (with the lens L2) and reflected (by a planar mirror located in the refocusing chamber) into the experimental chamber.    
In the monochromator, CG stands for ``classical geometry'' configuration, used for high-energy resolution measurements, while OPG represents the ``off-plane geometry'' configuration, used for high-temporal resolution measurements (see Fig.\ref{fig2} for details about the two configurations). 
\label{fig1}}
\end{figure}

The CITIUS laser system and the beamline are schematically shown in Fig.\ref{fig1}. 

\begin{figure}
\centering
{\resizebox{0.48\textwidth}{!}{ \includegraphics{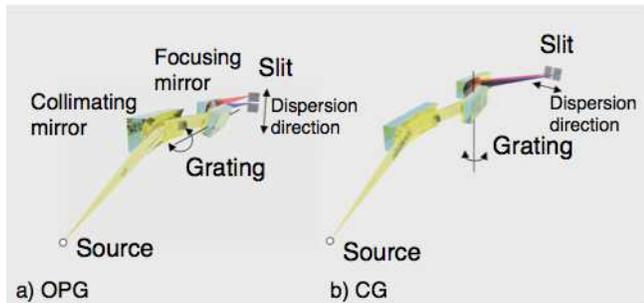}}} 
\caption{Light paths inside the monochromator: a) off plane geometry (OPG); b) classical geometry (CG). 
\label{fig2}}
\end{figure}

The laser system comprises a commercial amplifier “Legend Elite Duo” and an optical parametric amplifier “OPerA” (OPA), both produced by Coherent Inc. The amplifier is seeded by a Micra oscillator (wavelength: 800 nm, spectral bandwidth about 100 nm, power about 380 mW) and includes two amplification stages: a regenerative amplifier (EVO 30), pumped by an Evolution 30 laser (frequency-doubled Q-Switched Nd:YLF laser), and a single-pass amplifier (EVO HE), pumped by an Evolution HE laser (frequency-doubled Q-Switched Nd:YLF laser). The system can be operated at four different repetition rates: 100 Hz, 1 kHz, 5 kHz and 10 kHz. At 5 kHz, which is the currently adopted repetition rate, it generates pulses carrying about 3.1 mJ, with a duration of about 35 fs, centered at 805 nm. Two thirds of the energy is used for generating XUV pulses through HHG, one third as a pump for the OPA, or directly in combination with the HHG beam for pump-probe experiments. The OPA produces tunable radiation in the range between 0.4 eV and 5.6 eV, with variable energy per pulse (from few to hundreds of microjoules).      
\\
The beamline includes a high-vacuum section, through which the XUV beam propagates, and a section in air, used to transport the IR-UV light generated by the OPA. The XUV part comprises a HHG generation chamber and a monochromator. The laser is focused in the generation chamber (using the lens L1, see Fig. \ref{fig1}), where it interacts with a noble gas of choice (Ar or Ne, for the reported experiments) contained in a cell, generating the high-order harmonics. The harmonic generation efficiency can be optimized by varying the laser intensity, the gas pressure (typically, of the order of $10^{-3}$ mbar) and the relative position of the laser focus within the cell.
\\
The monochromator, whose scheme is shown in Fig. \ref{fig2}, was developed by the C.N.R.-I.F.N. Padova (Italy) and will be described in detail in a different paper \citep{mono}. It was designed to select one single harmonic, or a sub-band of it, in the spectral range between 250 and 25 nm (5-50 eV). However, the transmitted energy range could be further extended at the expense of efficiency, up to 100 eV. 
The CITIUS monochromator is superior to similar systems \cite{pole3,frasse}, in that it adopts a single-grating design with a double configuration \cite{pole1, pole4}: an off-plane geometry (OPG) for ultrafast response (few tens of femtoseconds) at the expense of energy resolution, and a classical geometry (CG) for high-energy resolution, albeit with long temporal response (several hundreds of femtoseconds). A first toroidal mirror collimates the light coming from the generation chamber and deflects it onto a planar (OPG or CG) diffraction grating. The diffracted light is then refocused by a second toroidal mirror onto the exit slit. The characteristics of the OPG and CG gratings are reported in Table \ref{table1}.

\begin{table}[ht]
\caption{Characteristics of the monochromator gratings. The energy resolution has been calculated on a 100-$\mu$m-wide exit slit.}
\centering 
\begin{tabular}{c c c c}
\hline\hline
Off-plane geometry &  & \\[0.5ex]
\hline
Grating 1 (groove density: 200gr/mm) &   \\
Spectral region & 100-250 nm (12-5 eV) \\
Energy resolution & 0.1 eV @ 10 eV  \\
Grating 2 (groove density: 400gr/mm) &   \\
Spectral region & 27-100 nm (45-12 eV) \\
Energy resolution & 0.2 eV @ 20 eV  \\
Grating 3 (groove density: 600gr/mm) &   \\
Spectral region & 12-40 nm (100-30 eV) \\
Energy resolution & 0.5 eV @ 40 eV  \\
\hline\hline
Classical geometry &  & \\[0.5ex]
\hline
Grating 1 (groove density: 300gr/mm) &   \\
Spectral region & 80-250 nm (12-5 eV) \\
Energy resolution & 10 meV @ 10 eV  \\
Grating 2 (groove density: 600gr/mm) &   \\
Spectral region & 50-100 nm (25-12 eV) \\
Energy resolution & 20 meV @ 20 eV  \\
Grating 3 (groove density: 1200gr/mm) &   \\
Spectral region & 30-60 nm (41-21 eV) \\
Energy resolution & 40 meV @ 40 eV  \\
\end{tabular}
\label{table1}
\end{table}

The branch transporting the OPA/IR beam includes a delay line, allowing one to control the optical path difference between the pump and probe beams with sub-micron precision. The OPA/IR beam is refocused by a lens (L2 in Fig. \ref{fig1}) and intercepted by a flat mirror (BK7, high-reflectivity, replaceable according to the specific wavelength in use) hosted in the recombination chamber.  At the exit of the monochromator, the XUV beam passes through a 200 nm thick Al filter (to stop the residual IR seed laser) and enters the recombination chamber. The chamber hosts a toroidal mirror (Si:Au coated, with 1200 mm focal length) to refocus the monochromatized XUV light at the sample position. The mirror sits in a two-axes motorized mount, to guarantee (with micrometric precision) a perfect spatial overlap between the pump and probe beams onto the sample. 
The pump and probe beams impinge onto the sample in an almost collinear geometry in order to minimize the temporal resolution spread due to the lateral spots’ dimensions. The HHG spot has been measured to be about 150 $\mu$m (FWHM) at the sample position, while the OPA beam focusing is adjusted in order to obtain a spot approximately twice as big. This ensures a homogeneous excitation of the probed area. The energy density of the IR-UV at the interaction point has been estimated to be of the order of 10 mJ/$\text{cm}^2$.  

\section{HHG performance}

\begin{figure}
\centering
{\resizebox{0.48\textwidth}{!}{ \includegraphics{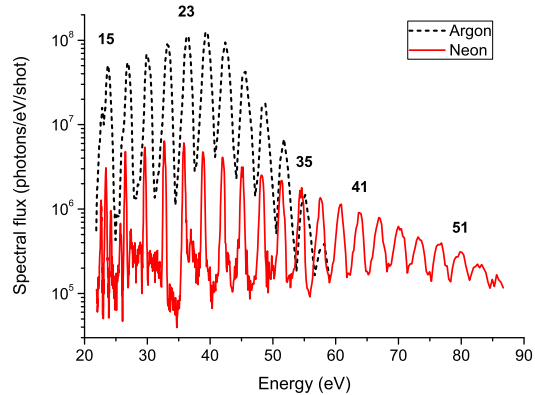}}} 
\caption{Harmonic spectra generated in Ar and Ne using laser pulses with energy $\simeq$ 2 mJ, duration $40\pm 2$fs (after transport to the generation chamber), wavelength = 805 nm and repetition rate = 5 kHz. Some of the harmonics have been labelled (see also Table \ref{table2}). The spectra have been acquired with a NIST-calibrated Al photodiode, placed at the exit of the monochromator. The monochromator slits aperture was set to 15 $\mu$m, in order to get a spectral resolution of about 0.05nm. 
\label{fig3}}
\end{figure}

Two typical spectra, obtained using Ar and Ne for harmonic generation, are shown in Fig. \ref{fig3}.

In agreement with existing literature, use of Ar produced harmonics in the photon energy range 20-54 eV, with a peak of about $1 \cdot 10^8$ photons/pulse at about 40 eV (25th harmonic). Using Ne, the plateau region moves towards higher energies, but with significantly reduced efficiency. In table 2 the generation efficiencies for the two gases at different harmonic orders are reported. 

\begin{table}[ht]
\caption{Generation efficiency after the monochromator with Ar and Ne. Selected harmonic orders are reported. In particular, the 35th (54.25 eV) and 41st (63.55 eV) harmonics generated with Ne, which correspond to the $M_{2,3}$ edges of Fe and Ni, respectively.}
\centering 
\begin{tabular}{c c c c c c c}
\hline\hline

\multicolumn{2}{c} {}  & $15^{th}$ & $23^{rd}$ & $35^{th}$ & $41^{th}$ & $51^{th}$\\
\multicolumn{2}{c}{} & ph/pulse  & ph/pulse & ph/pulse  &ph/pulse  &ph/pulse \\[0.5ex]
\hline
Ar &  & $3.4 \cdot 10^7$ & $1.1 \cdot 10^8$  &$2.2 \cdot 10^6$ &negligible&negligible\\
Ne &  & $7.1 \cdot 10^5$ & $2.2 \cdot 10^6$ &$1.5 \cdot 10^6$ &$1.1 \cdot 10^6$ &$5.6 \cdot 10^5$ \\ [1ex]
\hline
\end{tabular}
\label{table2}
\end{table}
In the following two sections, we report the results of two pump-probe experiments carried out to characterize the source performance. The experiment described in Section \ref{chara} aimed at measuring
the duration of harmonic pulses transmitted by the monochromator. The experiment described in Section \ref{magne} aimed at demonstrating the possibility of using CITIUS for studies of ultra-fast dynamics. 

\subsection{Characterization of harmonic pulse duration}
\label{chara}

\begin{figure}
\centering
{\resizebox{0.48\textwidth}{!}{ \includegraphics{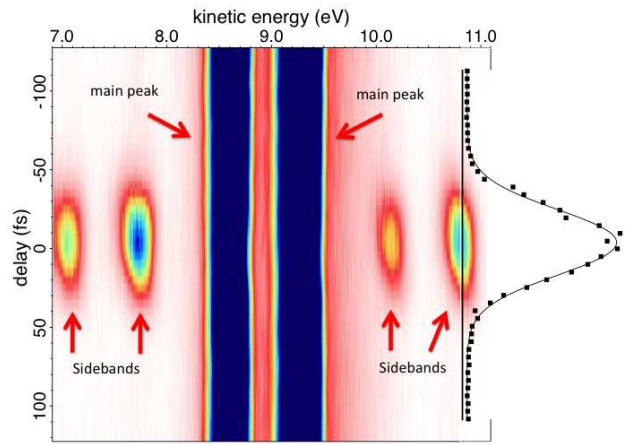}}} 
\caption{Kr 4p photo-electron spectra obtained by overlapping the XUV (pump) and IR (probe) beams. In this case, the XUV beam (15th harmonic of the fundamental) was monochromatized by the G200 grating (OPG geometry, see Table 1). The black dots represent the integrated intensity of the sideband at 10.8 eV, as a function of the pump-probe delay; the solid line is a Gaussian fit. The pulse duration is obtained by deconvolving the contribution of the IR pulse from the measured profile.
\label{fig4}}
\end{figure}

The pulse duration of HHG XUV pulses has been measured by means of an experiment based on LAPE (Laser Assisted Photoelectric Effect). This technique, described in detail in \cite{glove,toma}, is based on the cross correlation of XUV and IR photons. In essence, the primary photo-electron spectrum resulting from XUV ionization of a target gas (Kr, in the case under scrutiny) is modified if the ionization takes place in the presence of a high-intensity laser pulse: the absorption and emission of laser photons gives rise to so-called sidebands in the photoelectron spectrum. The amplitude of the sidebands as a function of the delay between the XUV and IR pulses provides a cross-correlation signal from which one can extract the duration of XUV pulses.  
\\
Our detection set-up was based on a hemispherical electron spectrometer (VG Scienta, SES-200) \cite{ulla}. The choice of Kr for the LAPE experiment was mainly dictated by instrumental reasons. The 15th harmonic (23.22 eV) has sufficient energy to ionize the Kr 4p levels and is efficiently transmitted by all monochromator gratings in the OPG configuration, with high-enough resolving power to fully resolve the $4p_{3/2}$ and $4p_{1/2}$ photo-electron lines. Additionally, the high photoionization cross section  of Kr 4p electrons grants fast acquisition times.
The laser pulse duration was carefully measured using a commercial autocorrelator and set at its minimum value, i.e., about $40\pm2$ fs after transport in the experimental chamber. The IR pulse was attenuated down to 80-100 mW in order to avoid saturation effects. 
Figure \ref{fig4} shows a series of Kr 4p photo-electron spectra (horizontal axis), as a function of the delay between XUV and IR pulses (vertical axis). The measurements have been performed using the 
G200 grating (see Table \ref{table1}). The two main photoelectron peaks, generated by the $4p_{3/2}$ and $4p_{1/2}$ states, have kinetic energies of 8.6 and 9.3 eV respectively. The presence of these peaks is due to a single-photon photoionization process and does not depend on the presence of the IR pulse. When the two pulses are temporally overlapped, two sidebands appear, corresponding to absorption and emission of one IR photon: these sidebands are at 7.8 and 10.8 eV for the $4p_{3/2}$ state, and at 7.1 and 10.1 eV for the $4p_{1/2}$ state. The intensity of each sideband as a function of the delay is the convolution of the XUV and IR pulses. The duration of the XUV pulse is obtained by deconvolving the IR signal from the Gaussian fit of the measured intensity. The black dots represent the temporal profile of the sideband at 10.8 eV; the solid line is a Gaussian fit. 
 
The extracted pulse durations are reported in Table \ref{table3}. Results clearly demonstrate the possibility to maintain a temporal resolution of few tens of femtoseconds after monochromatization using the OPG configuration.

\begin{table}[ht]
\caption{Measured pulse durations of the 15th harmonic (23.2 eV), transmitted by OPG gratings.}
\centering 
\begin{tabular}{c c c}
\hline\hline
Grating& Pulse duration (fs)\\[0.5ex]
\hline
G200 & 23 $\pm$ 1.2 \\
G400 & 35 $\pm$ 1.7 \\
G600 & 35 $\pm$ 1.9 \\[1ex]
\hline
\end{tabular}
\label{table3}
\end{table}

\subsection{Ultrafast demagnetization of permalloy}
\label{magne}

\begin{figure}
\centering
{\resizebox{0.50\textwidth}{!}{ \includegraphics{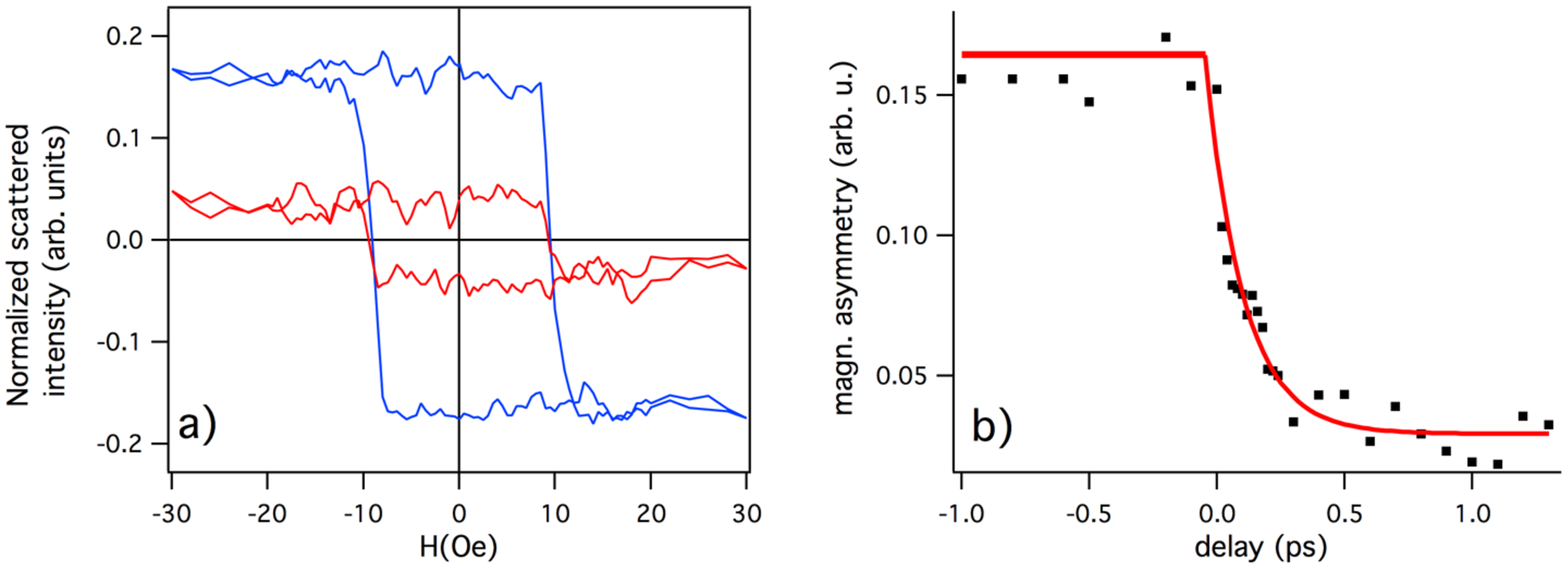}}} 
\caption{a) Hysteresis loops measured at the 35th harmonic of the fundamental ($M_{2,3}$ thresholds of transition metals) for HHG pulse arriving 1 ps before (blue line) and 1 ps after (red line) the IR pump pulse. b) Pump-probe delay dependency of the magnetic asymmetry ratio at saturation (dots) and exponential decay fit. The magnetization quenching is about 80 \% after 0.5 ps and the characteristic decay time is 0.21 $\pm$ 0.02 ps.
\label{fig5}}
\end{figure}

In this section, we report the results of a pump-probe experiment demonstrating the possibility of using CITIUS to study the ultra-fast dynamics of selected chemical species, with a temporal resolution of few tens of femtoseconds. 
\\ 
The spectral range covered by CITIUS includes the $M_{2,3}$ thresholds of transition metals (54.25 eV, i.e., the 35th harmonic of the fundamental, see Table \ref{table2}), making CITIUS an appealing source for the study of magnetic materials. As a proof-of-principle application, we measured the demagnetization of a 100 mm thick permalloy film ($\text{Fe}_{20}\text{Ni}_{80}$), using X-ray resonant magnetic scattering \cite{gibbs,refsXMRS2} in pump-probe configuration. 
\\
The experiment was performed using the IRMA reflectometer \cite{IRMA}. The instrument permits to precisely adjust the angles between the sample surface and the incoming/reflected XUV beam. Three photodiodes and one channel electron multiplier are mounted on the detector arm. By means of a horseshoe electromagnet, it is possible to apply a variable magnetic field up to $\pm 1500$ Oe, parallel to the sample surface. The scattering experiment was conducted in a transverse MOKE-like configuration \cite{ref T-MOKE}. In this configuration, using p-polarized XUV light, one is sensitive to the magnetization component perpendicular to the scattering plane. Moreover, choosing the incoming/scattering angle close to the Brewster condition (about 45 deg at 54 eV), the magnetization contrast is maximized with respect to the (strongly suppressed) non-magnetic scattered background, thus improving the signal-to-noise ratio of the measurement. 
\\
We have measured the magnetic hysteresis loop at the Fe $M_{2,3}$ edge, as a function of the delay between the transmitted IR beam (pump) and the XUV pulse (probe). Figure \ref{fig5}a shows the hysteresis loop when the XUV pulse arrives on the sample 1 ps before (blue curve) and 1 ps after (red curve) the IR pulse. The plots are normalized to the average scattered intensity. The magnetic asymmetry ratio is defined as $(I_+-I_-)/(I_++I_-)$, where $I_+$ and $I_-$ denote the intensities of the reflected light as the externally applied magnetic field is reversed in direction. In Figure \ref{fig5}b, the asymmetry ratio at saturation is plotted as a function of the pump-probe delay. Data show the ultrafast demagnetization of the Fe atoms of the sample. Almost 80 \% of the magnetization is dissipated 0.5 ps after the sample has been illuminated with the IR pulse. The data have been fitted with an exponential decay curve, indicating a characteristic demagnetization time of 0.21 $\pm$ 0.02 ps. 
Despite the modest scientific interest of this particular sample, which has been already investigated in detail by other groups \cite{REFbeaurepaire}, these results demonstrate that our apparatus is capable of measuring the ultrafast magnetization dynamics of selected chemical species, with a resolution of few tens of femtoseconds. 

\section{Conclusions and future developments}

We have presented the characteristics of CITIUS, a new IR-XUV, femtosecond, Italo-Slovenian light source, based on laser-induced high-harmonic generation in gas. As a next step, we will carry out the commissioning of the experimental equipment for time resolved photoemission investigations. The apparatus consists of two vacuum chambers equipped with two different electron spectrometers: an Angle-Resolved Time of Flight (VG-Scienta ARToF 10k) for pump-probe experiments and an hemispherical spectrometer (VG-Scienta R3000), which will be used in combination with a monochromatized twin anode X-ray source for conventional X-ray photoemission spectroscopy measurements. A preparation chamber, equipped with evaporators and standard sample preparation tools, such as sample heating, ion sputtering and LEED, will connect the two chambers. The sample will be hosted on a cryomanipulator (able to reach a temperature below 10 K), operated with liquid helium, with five degrees of freedom. 
\\
Starting from 2014, the source will be open to external users, interested in developing time-resolved applications in different domains of both fundamental and applied science. Envisaged experiments will be in particular related to photophysics and photochemistry, including materials science, catalysis, magnetism and biochemistry. Part of the activity will be carried out in close connection with that developed at the Low-Density Matter beamline of the FERMI free-electron laser. 

\section{Acknowledgements}

The CITIUS project is funded by the program for cross-border cooperation between Italy and Slovenia 2007-2013. C. Grazioli gratefully acknowledges financial support through the TALENTS Programme (7th Framework Programme, Specific Programme: PEOPLE - Marie Curie Actions - COFUND). 
We aknowledge the Department of Physics and Astronomy of Uppsala University for providing the VG-Scienta SES200, in the framework of the collaboration with CNR-IOM (TASC laboratory, Gas Phase beamline, Trieste). We also thank Carla Puglia and Monica de Simone for technical assistance in setting up the photoemission experimental set-up, and Sandra Gardonio for valuable suggestions about the source desing. G.D.N. is particularly thankful to Gvido Bratina for technical and moral support during the preparation and implementation of the project CITIUS.

\end{document}